\documentstyle[epsf]{l-aa}
\begin{document}

\thesaurus{08.16.7}

\title{Pulsed Optical Emission from Geminga
\thanks{Based on Observations taken at ESO, La Silla, Chile and SAO,
Nizhnij Arhyz, Russia }}

\author{A. Shearer \inst{1}
A. Golden \inst{2}, S. Harfst \inst{2}, R. Butler \inst{2}, R. M. Redfern \inst{2}, C. M. M. O'Sullivan \inst{2},
G. M. Beskin \inst{3}, S. I. Neizvestny \inst{3},  V. V. Neustroev \inst{3}, V. L. Plokhotnichenko \inst{3},
M. Cullum \inst{4}, A. Danks \inst{5}}
\offprints{A Shearer, andy.shearer@ucg.ie}

\institute{ Information Technology Centre, University of Galway, Galway, Ireland
\and Department of Physics, University of Galway, Galway, Ireland 
\and SAO, Nizhnij Arhyz, Karachai-Cherkessia, Russia
\and European Southern Observatory, Garching-bei-M\"{u}nchen, Germany
\and STX/Goddard Space Flight Centre, Greenbank, Maryland, USA}

\date{Received ...; Accepted ...}
\maketitle
\markboth{Shearer, Golden et al, Pulsed Optical Emission from Geminga}{}
\begin{abstract}

We present optical data which shows that G", the optical counterpart of
the $\gamma$-ray pulsar Geminga, pulses in B with a period of 0.237
seconds. The similarity between the optical pulse shape and the
$\gamma$-ray light curve indicates that a large fraction of the optical
emission is non-thermal in origin - contrary to recent suggestions based upon the
total optical flux. The derived magnitude of the pulsed emission is $m_B
= 26.0 \pm 0.4$. Whilst it is not possible to give an accurate figure
for the pulsed fraction (due to variations in the sky background) we can
give an upper limit of $m_B \approx 27 $ for the unpulsed fraction.
\end{abstract}


\section{Introduction}

The nature of the bright $\gamma$-ray source Geminga remained elusive
from the first observations using SAS-B (\cite{fich75}) until its
recognition as a pulsar with a period of 0.237 seconds in $\gamma$ rays
(\cite{bert92} \cite{big92}) and in X-rays (\cite{hal92}).  Based upon
colour considerations an optical candidate was proposed, G" with a m$_V$
of 25.5 (\cite{halp88}). This star had a measurable proper motion
(\cite{big93}) indicating a probable distance of about 100 pc and
thereby making a probable association with a neutron star.  Subsequent
Hubble Space Telescope observations have given a distance based upon
parallax of $159^{+59}_{-34}$ pc (\cite{car96}). 

Optical observations in B showed Geminga to be fainter than 26th
magnitude (\cite{big88}) - a result confirmed by HST observations
(\cite{big97}). In V Geminga is brighter at 25.4.  This aspect of the
spectrum has been explained by a proton cyclotron feature causing either
preferential emission in V or absorption in B and I (\cite{big96})
superimposed on a thermal continuum.  However, re-analysis of the
EUVE and ROSAT datasets highlight an error in this earlier work,
indicating that the thermal continuum would not be expected to dominate
in the optical regime, based on the observed flux (\cite{hal96}).
Such an apparent absorption feature has
been previosuly observed in the Crab spectrum (\cite{nas96}) 
although not confirmed
by other observations (\cite{kom96}). Recent spectral studies
of Geminga (\cite{mar98}) show a continuous power-law from 3700 to 8000
(\AA) with no such features consequently indicating that a
predominantly magnetospheric origin is preferred over a thermal one. It
should be noted that these spectroscopic studies were at the limit of the
observational capabilities of the Keck and with a low signal-to-noise
ratio. 

Of crucial importance to the understanding of neutron star structure is
the stellar radius. This can in principle be inferred once the distance and the
black-body contribution has been measured (\cite{wal97}). However
determining the black-body component of an isolated neutron star is
complicated by magnetospheric and possible atmospheric effects
(\cite{pav96}). As Geminga is very nearby it is a prime candidate for
measuring the thermal component - crucial to this will be the removal of
the magenetospheric component of its emission. This is possible by
determining what contribution of the optical emission is pulsed and
whether it 'follows' the hard (magnetospheric) or soft (presumed
thermal) X-ray emission profile. The faintness of the optical
counterpart has precluded time-resolved observations using conventional
photometers.  However by using 2-d photon counting detectors,  the
required astrometric analysis can be carried out off-line.  Consequently
photon arrival times can be measured from a reduced (seeing optimised)
aperture diaphram.

\section{Observations}

Observations were made on 25th and 26th February 1995 using the 3.55m
New Technology Telescope (NTT) at La Silla. Follow up observations were
taken in January 1996, using the 6m telescope (BTA) of the Special
Astrophysical Observatory  over three nights.  Two MAMA detectors were
used; one a B extended S-20 (\cite{tim86}) and the other a bialkali (\cite{cull90})
photocathode. By using the UCG TRIFFID camera (\cite{red93}) to record the data.
The arrival time and position of each photon was recorded to a precision
of 1 $\mu$second and 25 microns.  The spatial resolution was equivalent
to 0".13 on the NTT and 0".25 on the BTA.  Absolute timing was achieved
using a combination of a GPS receiver, which gave UTC to a precision of
400nsec every 10 seconds, and an ovened 10MHz crystal which was accurate
to $<$ 1 $\mu$second per 10 second interval. On each night the Crab
pulsar was observed for calibration purposes.    Using a Crab timing
ephemeris (\cite{lyne96}) the barycentric phase of the Crab pulse was determined;
phase was maintained to within 10 $\mu$seconds over the whole period. 
Table 1 shows a log of the observations. 

\begin{table*}
\begin{center}
\caption[]{Summary of Observations}
\label{obs}

\begin{tabular}{ccccccc}
\hline\noalign{\smallskip}
 Date          &  UTC    & Duration & Detector & Telescope & Filter  & Seeing\\
               & \null   & (s)      & \null    & \null     & \null   & ($\arcsec$)     \\ 
\noalign{\smallskip}
\hline\noalign{\smallskip}
1995 Feb 26    & 01:14:37 & 4580 & GSFC & NTT &  V & 1.3 \\
1995 Feb 26    & 02:36:37 & 4387 & GSFC & NTT &  V & 1.4 \\
1995 Feb 26    & 03:50:49 & 3662 & GSFC & NTT &  V & 1.4 \\
1995 Feb 27    & 01:58:20 & 788  & ESO  & NTT &  B & 1.3 \\
1995 Feb 27    & 02:42:23 & 2096 & ESO  & NTT &  B & 1.2 \\
1995 Feb 27    & 03:19:28 & 3000 & ESO  & NTT &  B & 1.7 \\
1996 Jan 12    & 18:07:15 & 4397 & ESO  & BTA &  B & 1.6 \\
1996 Jan 12    & 19:21:14 & 6409 & ESO  & BTA &  B & 1.5 \\
1996 Jan 12    & 21:36:04 & 884  & ESO  & BTA &  V & 1.5 \\
1996 Jan 12    & 21:52:02 & 413  & ESO  & BTA &  V & 1.5 \\
1996 Jan 12    & 22:23:56 & 2914 & ESO  & BTA &  V & 1.3 \\
1996 Jan 12    & 23:13:26 & 2618 & ESO  & BTA &  V & 1.4 \\
1996 Jan 13    & 19:28:34 & 7509 & ESO  & BTA &  B & 2.2 \\
1996 Jan 14    & 16:59:25 & 8182 & ESO  & BTA &  B & 1.2 \\
1996 Jan 14    & 19:18:44 & 2810 & ESO  & BTA &  B & 1.2 \\
\noalign{\smallskip}
\hline
\end{tabular}
\end{center}
\end{table*}

Photon positions were binned to produce an image after each exposure was
made.  By using the TRIFFID image processing software, the images could
be marginally improved by removing the effects of telescope
movement (\cite{she96}).  These images were compared with HST/WFPC2 archival
images to determine the position of Geminga at these epochs. 
After coaddition of all the B and V images from January 1996, a faint
star could be seen at the expected position of Geminga. No such object
could be seen in the February 1995 data. The reason for this was two
fold: firstly the exposure time-telescope aperture product was 5 times
greater in 1996 compared to 1995 and secondly the flat-fields were
deeper in the later observations.

Once the position of Geminga was established, the photons times were
extracted from a window, centred on Geminga,  with a diameter
corresponding to the average seeing widths for each exposure.  This was
chosen to maximise the signal to noise ratio.  These extracted times
were then translated to the solar system barycentre using the JPL DE200
ephemeris. The Geminga arrival times were folded in phase using the
EGRET ephemeris (\cite{matt96}) for each colour and for each observing
run. 
\begin{figure} 
\epsfysize 3.3truein
\epsffile{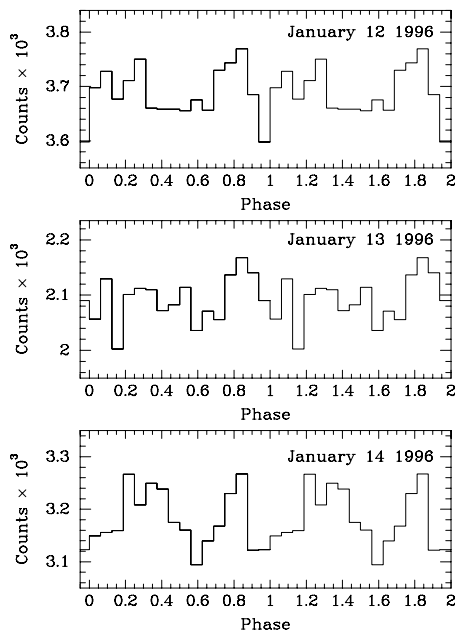}
\caption{
Phase plot for the three individual nights observed in January 1996. Two
phases are shown for clarity. \label{fig1} }
\label{Individual Phase Plots}
\end{figure}

\begin{figure}  
\epsfysize 3.3truein
\epsffile{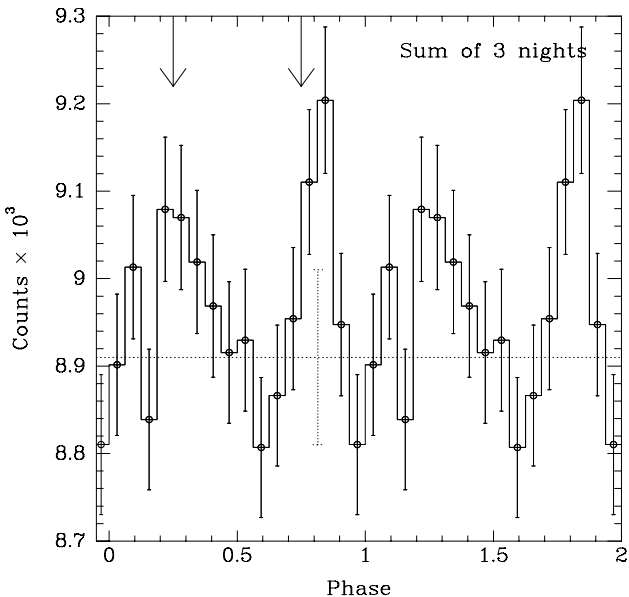} 
\caption{  Total phase plot for
January 1996 in B. The error bars represent the Poissonian fluctuations
of the original data set. Also marked are the phases of the peaks of the
EGRET light curve. The dotted line indicates the background level with
its associated error based upon counting statistics ($\pm 25$) and
systematic errors including flat field errors ($\pm 75$).  }
\label{Figure 2}
\end{figure}

Figure 1 shows the light curve in B for each night of January 1996 and
Figure 2 shows the combined light curve. Figure 3 shows the time
resolved images in B for 4 phase bands as indicated. Clearly Geminga can
be seen during the two 'on' phases; star G is to the top right. We note
that background fluctuations are consistent with the expectations of
Poissonian statistics. Table 2 shows the determined pulsed magnitude or
1 sigma upper limits where appropriate.  This type of light curve is
best analysed using the $Z^2_n$ statistic (\cite{buc89}).  Only the
January 1996 B data shows significant pulsations - the $Z^2_2$ statistic
for this data set (of 20.6) has a significance of 99.96\%. In the
January 1996 V data a weak signal, with a similar form, can also be
observed. 

\begin{table}
\caption[] {Fluxes}
\begin{flushleft}
\begin{tabular}{ccc}
\hline\noalign{\smallskip}
Data Set & Pulsed 3 $\sigma$ Upper Limits & Flux $\mu$Jy\\
\noalign{\smallskip}
\hline\noalign{\smallskip}
95V & 24.0 &  0.95 \\
95B & 23.8 & 1.3  \\
96V & 25.5 &  0.24 \\
96B & 26.0 $\pm 0.4 $ & 0.17 {$^{+0.07}_{-0.05}$} \\
\noalign{\smallskip}
\hline
\end{tabular}
\end{flushleft}
\end{table}

This can be understood from the length of the V observation being about a fifth of
the B, the sky brightness in V was about 1.5 magnitudes higher than in B
and the detector used in 1996 had a bi-alkali photocathode with a B DQE
higher than V (\cite{cull90}). No significant signal was observed in
February 1995 - mainly due to the smaller telescope aperture. Given
these considerations the upper limits of our data are consistent with
the level of pulsations remaining constant. In order to estimate the
pulsed fraction the background level had to measured.  For each data set
the background was determined as the mean of  the signal in a small
annulus, of radius 2".5 and width 0".25 around the position of Geminga. 
The magnitude calibration was achieved using the star G {(\cite{big88})}
which was always in the field of view. Our measured pulse fraction is
consistent with 100\%, but, we should stress that there is a large error
in this value, due to uncertainties in measuring the sky background, as
seen in figure 2. We can however give a $1 \sigma$ upper limit to the
unpulsed component of $30\%$.

\begin{figure}  
\epsfysize 3.3truein
\epsffile{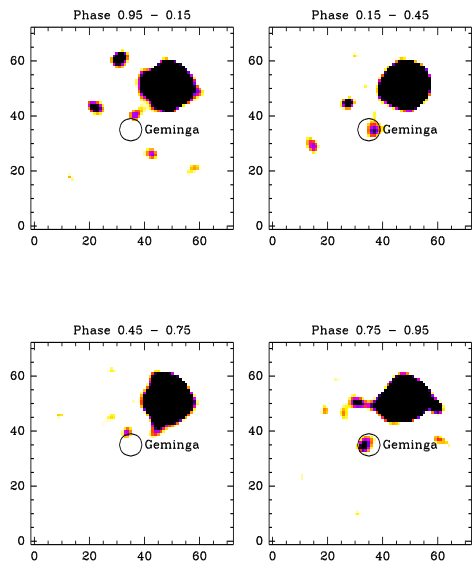} 
\caption{  Phase resolved images of
Geminga. The pixel scale is 0".22 / pixel. Star G is to the top right in
all images and Geminga is marked. }
\label{Figure 3}
\end{figure}

\section {Discussion}

Our results agree qualitatively with the previously observed optical
flux, but the form of the optical light curve resembles the $\gamma$ and
hard X-ray rather than the soft X-ray signature, implying a
magnetospheric origin.  In $\gamma$-rays the spectrum is double peaked
with maxima at phases 0.25 and 0.75 (\cite{matt96}). The soft X-ray is
characterized by a sinusoidal modulation consistent with thermal
emission with two temperatures ($5.2$ $10^5$ and $3.0$  $10^6$K) or from
a single thermal source and a power spectrum (\cite{hal93},
\cite{hal97}). EUVE satellite observations, albeit with low
significance, indicate that the extreme UV timing profile is similar in
shape to the soft X-ray light curve. Recent reported detections of radio
emission (\cite{kuz97}, \cite{mal97}) do not show a consistent pulse
profile pattern and Geminga's radio emission is still a topic of some
debate. However the phase agreement of the $\gamma$-ray light curve
with the radio profile of Kuz'min and Losovskii, using the EGRET
ephemeris, would suggest that peak 2 corresponds to direct polar
emission as reported in their paper. When this is combined with the
doubled pulsed optical, hard X-ray and $\gamma$-ray profile we can
attempt to constrain  models for the high energy emission. The high
energy pulses are about 0.5 phase apart which rules out a geometry
similar to the Crab pulsar. It would seem that Geminga has a magnetic
axis at about $90^o$  to the rotation axis, with the emission
sites situated close to the neutron star surface. Interestingly when we
fold our data, using the  Malofeev and Malov ephemeris, one of the
optical pulse peaks arrives  at phase zero - expected from most emission
models where phase zero  generally corresponds to the radio peak. What
is not clear is why the Malofeev and Malov pulse is delayed to phase
0.4.

From Figure 2 we can see that the signal shows two peaks with a phase
separation of $\approx$ 0.5. We can understand this in terms of the
extrapolation of the pulsed $\gamma$ emission to optical wavelengths,
which is valid for the Crab pulsar. $\gamma$-ray emission  from Geminga
has been suggested  to be variable (\cite{ram95}) in both total
intensity and spectral index.  Our optical observations are within the
spread of variation and wide dispersion from the higher energy points
(\cite{may94}). Halpern \& Wang's more recent analysis fitted the x-ray
and hard UV data to a black-body spectrum (T $\approx 6~~10^5$ K) with a
power law. The low energy extrapolation of their black-body fit would
produce an optical flux 2 magnitudes fainter than we observe, whilst the
power law would produce a flux  $>$ 5 magnitudes brighter. As with the
Crab pulsar this points to a probable flattening of the spectrum in the
UV region.  We should also note this extension from the X-ray to the
optical does not require a proton cyclotron feature (\cite{big96}) to
explain the V data but it might imply that there is preferential
absorption in I. This could be due to a number of processes including 
electron cyclotron resonance scattering or synchrotron self-absorption
(unlikely in this instance). Such a turnoff has been observed
spectroscopically in Geminga (\cite{mar98}) albeit with low
significance. As similar I attenuation is observed in both the Crab and
Vela pulsars' spectra we are begining to develop an understanding
whereby the high energy emission from these objects comes from a similar
process. Our results when taken in conjunction with the spectroscopy
studies favour an electron synchrotron origin for the radiation at least
up to the edge of the B band. Differences can be understood in terms of
the viewing geometry and path length over which the electrons are
radiated. Deep, phase resolved observations in V, R, I and H will be
crucial in determining whether there is a continuum infra-red turnover
or whether there is an absorption line.  Such observations, including a
determination of the pulse fraction, will adequately separate the
thermal and non-thermal components with important ramifications for
models of pulsar emission. Indeed, the unpulsed  components would
constrain Geminga's thermal continuum, which in conjunction with EUVE
and X-ray data and the known distance would provide definitive
estimates, rather than upper limits (\cite{wal97}), on the  neutron
star's size.

{\it  ESO and Goddard Space Flight Centre is thanked for the provision
of their MAMA detector.  Peter Sinclair and the ESO detector workshop at
La Silla are thanked for their invaluable help during the ESO
observations. We are also grateful to the 6-m telescope Program
Committee of the RAS for observing time allocation.  We thank the
engineers of SAO RAS, A.  Maksimov for help in equipment preparation for
the observations and the Director of SAO RAS Yu.  Balega for arranging
the observations.  This work was supported by the Russian Foundation of
Fundamental Research, State programme "Astronomy", Russian Ministry of
Science and Technical Politics, and the Science-Educational Centre
"Cosmion".  The support of FORBAIRT, the Irish Research and Development
agency, is gratefully acknowledged.  Peter O' Kane of University of
Galway is thanked for technical assistance during the observations.
}

\end{document}